\documentclass[leqno]{siamltex}

\usepackage[all]{xy}
\usepackage{amsmath,amssymb,mathtools}

\usepackage[titletoc,title]{appendix}

\usepackage[hyphens]{url}
\usepackage{graphicx}
\usepackage{pgfplots}
\usepackage{float}

\usepackage{appendix}
 \usepackage{microtype} 

\usepackage{amsthm}
\theoremstyle{plain}\newtheorem{mythm}{Theorem}
\theoremstyle{definition}\newtheorem{mydef}{Definition}
\theoremstyle{remark}\newtheorem{myrem}{Remark}
\theoremstyle{definition}\newtheorem{myassum}{Assumption}

\newcommand{\Var}{\operatorname{Var}}

\usepackage{etoolbox}
\apptocmd{\thebibliography}{\raggedright}{}{}

\usepackage{tikz}
\usetikzlibrary{calc,shapes,decorations.markings,shapes.geometric}
\usetikzlibrary{arrows,positioning,backgrounds} 
\usetikzlibrary{plotmarks} 

\usetikzlibrary{topaths,intersections,calc,through,spy,fit}
\usepackage{ifthen}
\newcommand{\ifnonzero}[2]{
 \ifthenelse{\equal{#1}{0}}{}{#2}
}
\tikzstyle{glassware} = [fill=white]

\newcommand{\flask}[3][]{
 \begin{scope}[shift={(#2)},glassware,#1]
  \ifnonzero{#3}{
   \begin{scope}
    \clip[rounded corners]
         (-0.45,3) -- (-0.25,3) -- (-0.25,2) -- (-1.0,0) -- (1.0,0) -- (0.25,2) -- (0.25,3) -- (0.45,3);
    \filldraw
         [yscale=2.8] (-1.0,#3)
         [rounded corners] -- (-1.0,0) -- (1.0,0)
         [sharp corners] -- (1.0,#3) -- cycle;
   \end{scope}
  }
  \draw[rounded corners]
        (-0.45,3) -- (-0.25,3) -- (-0.25,2) -- (-1.0,0) -- (1.0,0) -- (0.25,2) -- (0.25,3) -- (0.45,3);
 \end{scope}
}

\tikzset{
    >=stealth',
    normrect/.style={
           rectangle,
           rounded corners=5mm,
           draw=black, very thick,
           text centered,
           align=center,
           minimum height=1.5cm, minimum width=3cm},
    normarr/.style={
           ->,
           thick,
           shorten <=2pt,
           shorten >=2pt},
    LBD/.style={
           rectangle,
           rounded corners=1.5mm,
          draw=red,
           text centered,
           align=center,
           minimum height=0.3cm, minimum width=0.3cm,
           fill=red!10},
      FUS/.style={
          rectangle,
           rounded corners=3mm,
           draw=blue,
           text centered,
           align=center,
           minimum width=0.6cm, minimum height=0.6cm,
           fill=blue!10}
}

\definecolor{prp}{RGB}{255,0,255}
\definecolor{lgr}{RGB}{200,200,200}

\title{Modeling, Simulating, and Parameter Fitting of Biochemical Kinetic Experiments}
\author{D. Goulet\footnotemark[2]}

\begin{document}

\maketitle

\renewcommand{\thefootnote}{\fnsymbol{footnote}}

\footnotetext[2]{goulet@rose-hulman.edu}

\begin{abstract}
In many chemical and biological applications, systems of differential equations containing unknown parameters are used to explain empirical observations and experimental data. The DEs are typically nonlinear and difficult to analyze, requiring numerical methods to approximate the solutions. Compounding this difficulty are the unknown parameters in the DE system, which must be given specific numerical values in order for simulations to be run.

Estrogen receptor protein dimerization is used as an example to demonstrate model construction, reduction, simulation, and parameter estimation. Mathematical, computational, and statistical methods are applied to empirical data to deduce kinetic parameter estimates and guide decisions regarding future experiments and modeling. The process demonstrated serves as a pedagogical example of quantitative methods being used to extract parameter values from biochemical data models.

\end{abstract}



\pagestyle{myheadings}
\thispagestyle{plain}
\markboth{D. Goulet}{Modeling, Simulating, and Parameter Fitting in Biochemistry}

\section{Introduction} 
\begin{figure}[htbp]
\centering
\begin{tikzpicture}
	\newlength\figureheight 
	\newlength\figurewidth
\node at (2.75,5.75) [normrect, fill=white] (T) {Theoretical\\ Model};
\node at (-1,4.125) [normrect, fill=white] (A) {Mathematical\\ Model};
\node at (6.5,4.125) [normrect, fill=white] (E) {Experimental\\ Model};
\node at (-1,0.25) [normrect, fill=white] (C) {Computational\\ Model};
\node at (2.75,1.875) [normrect, fill=white] (D) {Data\\ Model};

\path (T) edge[normarr, in=90, out=180] (A);
\path (T) edge[normarr, in=90, out=0] (E);
\path (A) edge[normarr, in=90, out=270] (C);
\path (C) edge[normarr, in=270, out=0] (D);
\path (E) edge[normarr, in=0, out=270] (D);
\path (D) edge[normarr, in=270, out=90] (T);
\path (D) edge[normarr, in=90, out=180] (C);

\begin{pgfonlayer}{background}
\draw[draw=black!20, fill = lgr!10]
(-2.75, -0.75) rectangle (4.5, 2.875);
\draw (3.625, -0.375) node{\footnotesize \color{black!80}\emph{in silicio}};

\draw[draw=black!20, fill = lgr!10]
(4.75, -0.75) rectangle (8.25, 6.75);
\draw (7.375, -0.375) node{\footnotesize \color{black!80}\emph{in vitro}};

\draw[draw=black!20, fill = lgr!10]
(-2.75, 3.125) rectangle (4.5, 6.75);
\draw (-1.75, 6.375) node{\footnotesize \color{black!80}\emph{in mente}};

\end{pgfonlayer}

\end{tikzpicture}
\caption[Modeling Cycle]{The flow of information for the dimer exchange model.}
\label{fig: modelcycle}
\end{figure}
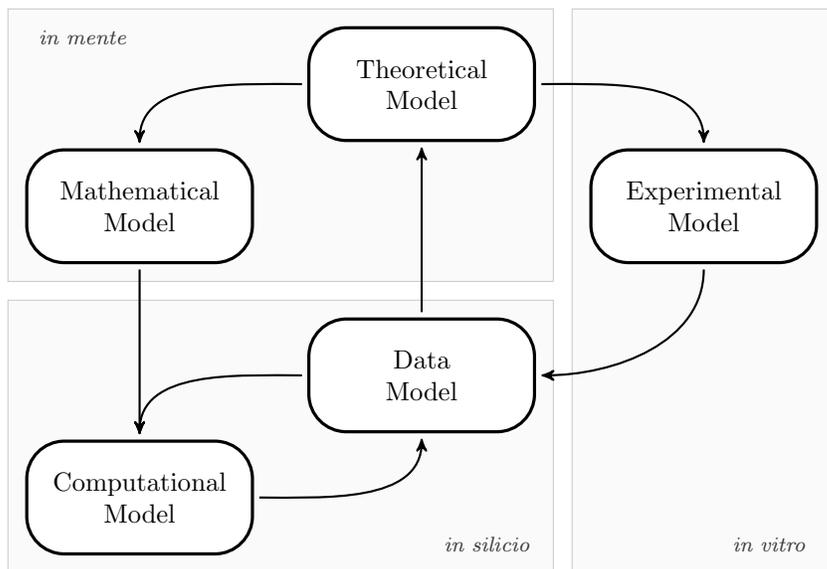

The empirical study of many current biological problems generates large and complex data sets. How best to use this data to generate and improve scientific hypotheses is a subject of great interest to biologists, mathematicians, statisticians, and computational scientists. Combining these various scientific and quantitative disciplines requires careful communication. Figure~\ref{fig: modelcycle} illustrates the flow of information for a typical problem in quantitative biology.

The modeling process described in the present work began with the formation of a scientific hypothesis based on laboratory experiments and intuition. This theory was used by biochemists to construct an experimental protocol and generate data. The same theory was used by mathematicians to develop a mathematical model, which was studied analytically and simulated computationally. Both models were intended to confirm or improve hypotheses. The theory and experiments are described in \S\ref{sec: biomodel} while the mathematical and computational models are described in \S\ref{sec: mathmodel} and \S\ref{sec: compmodel}, respectively.

For the biological model to feed back on theory, its output data needed to be analyzed. For the quantitative model to feed back on theory, its unknown parameter values needed to be estimated, so that meaningful simulations could be performed. Data modeling allowed empirical evidence to be combined with computational simulation as a means of generating parameter estimates and furthering biological theories. This process is described in \S\ref{sec: stats}.

As Figure~\ref{fig: modelcycle} suggests, the deduction of parameter estimates and confirmation of scientific hypotheses is not the end of the modeling process. Indeed, the data model answered some questions while raising others, prompting a new cycle of modeling which is now underway. Conclusions and future modeling directions are discussed in \S\ref{sec: conclusions}.

\section{Biochemical Modeling}
\label{sec: biomodel}
Cells contain a vast array of proteins serving a multitude of functions. One class of protein is called a \emph{receptor}. The specific type of receptor of interest is the estrogen receptor ER$\alpha$, found in various cell types, including human breast and ovary cells. This receptor is localized to the nuclear membrane and binds to estrogen molecules, ultimately leading to the expression of genes and the production of proteins. Because these receptors have been correlated to breast cancer, they are of interest to biochemists and drug developers \cite{Bieche:2001aa}.

Understanding the molecular structure of ER$\alpha$, and the mechanism by which it acts, can be approached on many experimental and analytical levels. Below a model is developed and analyzed to fit data gathered by biochemists attempting to understand the rates of formation of estrogen receptors \emph{in vitro} \cite{Brandt:1997aa}.

The laboratory protocol employed is known as the \emph{Dimer Exchange Assay}. It makes use of size exclusion chromatography to separate molecules of different sizes. For completeness, a brief description of this type of chromatography is given, followed by a more detailed description of the dimer exchange assay itself.

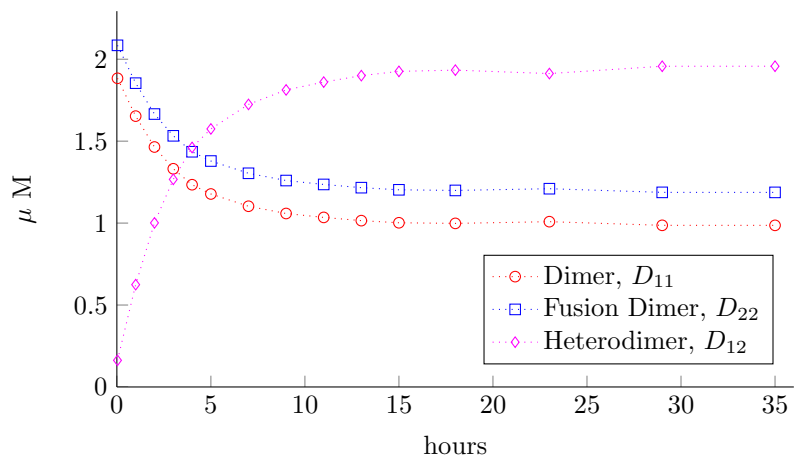
\begin{figure}[htbp]
	\centering
	\setlength\figureheight{5cm} 
	\setlength\figurewidth{9cm}
%
%
%
%

\definecolor{mycolor1}{rgb}{1,0,1}

\begin{tikzpicture}

\begin{axis}[%
width=\figurewidth,
height=\figureheight,
scale only axis,
xmin=0,
xmax=36,
xlabel={hours},
ymin=0,
ymax=2.29322177809321,
ylabel={$\mu\text{ M}$},
axis x line*=bottom,
axis y line*=left,
legend style={at={(0.97,0.06)},anchor=south east,draw=black,fill=white,legend cell align=left}
]
\addplot [
color=red,
dotted,
mark=o,
mark options={solid}
]
table[row sep=crcr]{
0.0333333333333333 1.88360533049696\\
1 1.65272682939082\\
2 1.4642963947004\\
3 1.33144387275514\\
4 1.23415578413048\\
5 1.17758068416964\\
7 1.10279495120608\\
9 1.05853706922122\\
11 1.03468981985722\\
13 1.01497696566523\\
15 1.00197713980928\\
18 0.99815835867658\\
23 1.00864866255774\\
29 0.986343078158561\\
35 0.986240444902865\\
};
\addlegendentry{Dimer, $D_{11}$};

\addplot [
color=blue,
dotted,
mark=square,
mark options={solid}
]
table[row sep=crcr]{
0.0333333333333333 2.08474707099383\\
1 1.85386856988769\\
2 1.66543813519727\\
3 1.53258561325201\\
4 1.43529752462736\\
5 1.37872242466651\\
7 1.30393669170295\\
9 1.25967880971809\\
11 1.2358315603541\\
13 1.2161187061621\\
15 1.20311888030615\\
18 1.19930009917345\\
23 1.20979040305462\\
29 1.18748481865543\\
35 1.18738218539974\\
};
\addlegendentry{Fusion Dimer, $D_{22}$};

\addplot [
color=mycolor1,
dotted,
mark=diamond,
mark options={solid}
]
table[row sep=crcr]{
0.0333333333333333 0.162517865219867\\
1 0.624274867432142\\
2 1.00113573681298\\
3 1.2668407807035\\
4 1.46141695795281\\
5 1.57456715787449\\
7 1.72413862380162\\
9 1.81265438777133\\
11 1.86034888649933\\
13 1.89977459488333\\
15 1.92577424659522\\
18 1.93341180886062\\
23 1.91243120109829\\
29 1.95704236989666\\
35 1.95724763640805\\
};
\addlegendentry{Heterodimer, $D_{12}$};

\end{axis}
\end{tikzpicture}%
	\caption[Dimer Exchange Data]{Data from a dimer exchange assay. See Table~\ref{tab: dimerexchangedata} in Appendix~\ref{app: data}.}
	\label{fig: dimerexchangedata}
\end{figure}

Size exclusion chromatography is a technique used to separate molecules of different sizes and quantify the concentrations of molecules as they change over time. It is an effective tool for understanding the rates at which biochemical reactions occur. Reacting molecules are placed on one end of a glass column filled with starchy beads. As the molecules diffuse and convect through the column, their interactions with the beads, which vary depending on the geometry of the molecule, cause the molecules to be separated by size. The data extracted from the chromatography column helps to reconstruct the time course of concentrations. A typical time course is illustrated in Figure~\ref{fig: dimerexchangedata}.

Models of the size-exclusion chromatography mechanism are not pursued here. But there has been activity in this area of theoretical and computational modeling~\cite{Felinger:2008aa, Lou:2004aa, Yu:2006aa}.

\begin{figure}[htbp]
\centering
\begin{tikzpicture}

\node at (1.5,1.3) [LBD] {};
\draw (1.5, 0.75) node{\footnotesize \color{black!80} LBD monomer$\strut$}; 

\node at (5.25,1.3) [FUS] {};
\draw (5.25, 0.75) node{\footnotesize \color{black!80} LBD/MBP fusion monomer};

\node at (-0.35,0) [LBD] {};
\node at (-0.15,0) [LBD] {};
\draw (-0.25, -0.55) node{\footnotesize \color{black!80} homodimer$\strut$};

\node at (3.2,0) [FUS] {};
\node at (3.55,0) [FUS] {};
\draw (3.375, -0.55) node{\footnotesize \color{black!80} fusion homodimer};

\node at (7.25,0) [FUS,draw=prp,fill=prp!10] {};
\node at (7.55,0) [LBD,draw=prp,fill=prp!10] {};
\draw (7.25, -0.55) node{\footnotesize \color{black!80} hererodimer$\strut$};

\end{tikzpicture}
\caption[Dimer Exchange]{The five proteins of the dimer exchange assay. Maltose Binding Protein is fused with the Ligand Binding Domain protein of ER$\alpha$. The two monomer types dimerize in three ways.}
\label{fig: dimershapes}
\end{figure}
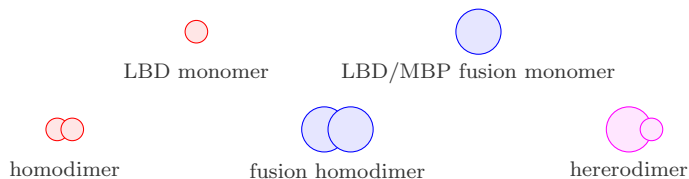

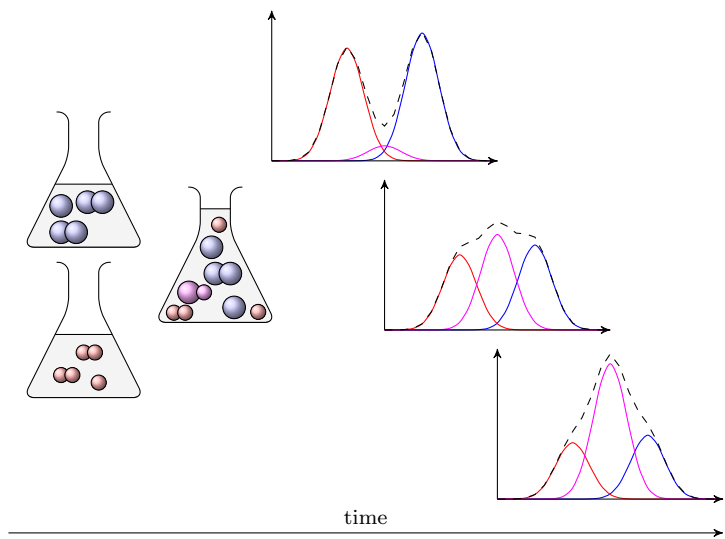
\begin{figure}[htbp]
\centering
\begin{tikzpicture}[node distance=1cm, auto]

\flask[xscale=0.8,yscale=0.6,fill=lgr!20]{0,1}{0.5};
\flask[xscale=0.8,yscale=0.6,fill=lgr!20]{0,3}{0.5};
\flask[xscale=0.8,yscale=0.6,fill=lgr!20]{1.75,2}{0.9};

\shadedraw [ball color= red!30] (-0.3,1.3) circle (0.1cm);
\shadedraw [ball color= red!30] (-0.15,1.3) circle (0.1cm);
\shadedraw [ball color= red!30] (0,1.6) circle (0.1cm);
\shadedraw [ball color= red!30] (0.15,1.6) circle (0.1cm);
\shadedraw [ball color= red!30] (0.2,1.2) circle (0.1cm);

\shadedraw [ball color= blue!20] (0.05,3.6) circle (0.15cm);
\shadedraw [ball color= blue!20] (0.25,3.6) circle (0.15cm);
\shadedraw [ball color= blue!20] (-0.3,3.2) circle (0.15cm);
\shadedraw [ball color= blue!20] (-0.1,3.2) circle (0.15cm);
\shadedraw [ball color= blue!20] (-0.3,3.55) circle (0.15cm);

\shadedraw [ball color= red!30] (1.8,3.3) circle (0.1cm);
\shadedraw [ball color= blue!20] (2,2.2) circle (0.15cm);
\shadedraw [ball color= blue!20] (1.75,2.65) circle (0.15cm);
\shadedraw [ball color= blue!20] (1.95,2.65) circle (0.15cm);
\shadedraw [ball color= prp!30] (1.4,2.4) circle (0.15cm);
\shadedraw [ball color= prp!30] (1.6,2.4) circle (0.1cm);
\shadedraw [ball color= red!30] (1.2,2.13) circle (0.1cm);
\shadedraw [ball color= red!30] (1.35,2.13) circle (0.1cm);
\shadedraw [ball color= red!30] (2.32,2.14) circle (0.1cm);
\shadedraw [ball color= blue!20] (1.7,3) circle (0.15cm);

\draw[->] (2.5,4.15) -- (2.5,6.15);
\draw[scale=1,domain=2.5:5.5,smooth,variable=\x,blue] plot ({\x},{1.7*exp(-10*(\x-4.5)^2)+4.15});
\draw[scale=1,domain=2.5:5.5,smooth,variable=\x,red] plot ({\x},{1.5*exp(-10*(\x-3.5)^2)+4.15});
\draw[scale=1,domain=2.5:5.5,smooth,variable=\x,prp] plot ({\x},{0.2*exp(-10*(\x-4)^2)+4.15});
\draw[scale=1,domain=2.5:5.5,dashed,variable=\x,black] plot ({\x},{1.7*exp(-10*(\x-4.5)^2)+ 1.5*exp(-10*(\x-3.5)^2)+0.2*exp(-10*(\x-4)^2)+4.15});
\draw[->] (2.5,4.15) -- (5.5,4.15);

\draw[->] (4,1.9) -- (4,3.9);
\draw[scale=1,domain=4:7,smooth,variable=\x,blue] plot ({\x},{1.13*exp(-10*(\x-6)^2)+1.9});
\draw[scale=1,domain=4:7,smooth,variable=\x,red] plot ({\x},{1*exp(-10*(\x-5)^2)+1.9});
\draw[scale=1,domain=4:7,smooth,variable=\x,prp] plot ({\x},{1.27*exp(-10*(\x-5.5)^2)+1.9});
\draw[scale=1,domain=4:7,dashed,variable=\x,black] plot ({\x},{1.13*exp(-10*(\x-6)^2)+ 1*exp(-10*(\x-5)^2)+1.27*exp(-10*(\x-5.5)^2)+1.9});
\draw[->] (4,1.9) -- (7,1.9);

\draw[->] (5.5,-0.35) -- (5.5,1.65);
\draw[scale=1,domain=5.5:8.5,smooth,variable=\x,blue] plot ({\x},{0.85*exp(-10*(\x-7.5)^2)-0.35});
\draw[scale=1,domain=5.5:8.5,smooth,variable=\x,red] plot ({\x},{0.75*exp(-10*(\x-6.5)^2)-0.35});
\draw[scale=1,domain=5.5:8.5,smooth,variable=\x,prp] plot ({\x},{1.8*exp(-10*(\x-7)^2)-0.35});
\draw[scale=1,domain=5.5:8.5,dashed,variable=\x,black] plot ({\x},{0.85*exp(-10*(\x-7.5)^2)+ 0.75*exp(-10*(\x-6.5)^2)+1.8*exp(-10*(\x-7)^2)-0.35});
\draw[->] (5.5,-0.35) -- (8.5,-0.35);

\draw[->] (-1,-0.8) -- node[above] {\footnotesize time} (8.5,-0.8) ;

\end{tikzpicture}
\caption[Dimer Exchange Sequence]{The sequence of events in the dimer exchange assay. Solutions of LBD and fusion homodimers  are prepared and allowed to equilibrate before being mixed. Free monomers are present in both mixed and unmixed solutions, but in relatively low concentrations. At regular time intervals, a sample is extracted from the mixture and injected into a high pressure liquid chromatography column. The areas of resulting chromatograms (black dotted curves) indicate the amount of protein in solution around the time of the injection. Peak fitting algorithms are applied to the chromatograms to determine the amounts of LBD dimer (left red curves), fusion dimer (right blue curves), and heterodimer (center purple curves) present in the effluent. Free monomer concentrations are below the noise threshold and can't be reliably estimated.}
\label{fig: exchangesequence}
\end{figure}

The dimer exchange assay is illustrated in Figure~\ref{fig: exchangesequence}. The ligand binding domain protein (LBD) of the estrogen receptor exists in monomeric form, but spontaneously dimerizes in solution. The fusion monomer is created by the addition of maltose binding protein to the LBD. This leads to the formation of two other dimer types, the fusion dimer and the heterodimer. See Figure~\ref{fig: dimershapes}.

It's not possible to produce solutions of free LBD monomers and observe the rate at which dimers form. A dilution of an equilibrated dimer/monomer solution would force concentrations away from equilibrium and allow kinetics to be observed. However, in the case of LBD estrogen protein, this equilibration occurs much too rapidly (on the order of two minutes) to be observed experimentally. To overcome this, the dimer exchange assay is performed. Once the solutions of homodimers are mixed, equilibrium is approached over roughly a 24 hour period. This allows sufficient time for a dimer concentration time series to be recorded.

The remainder of this article is presented as follows. In \S\ref{sec: mathmodel} estrogen receptor experiments are described and a mathematical model of the dimer exchange assay is constructed. The model is reduced using conservation laws and experimental observations. In \S\ref{sec: compmodel} computational methods for simulating and analyzing the model are described. Numerical simulations and optimization algorithms provide estimates of biochemical parameters. In \S\ref{sec: stats}, data analytic techniques are used to assess the accuracy of these estimates. In \S\ref{sec: conclusions} a general procedure for approaching similar biochemical problems is outlined, highlighting the importance of interaction between modelers and experimenters. Computational algorithms, experimental data, and related information are provided in the appendix.

\section{Mathematical Modeling}
\label{sec: mathmodel}
\subsection{Reaction Rate Laws}
\label{sec: rate laws}
Estrogen Receptors are dimers not monomers, \emph{i.e.}, made of two proteins not one. The two monomeric proteins are identical. Letting $D$ represent the dimer and $M$ represent the monomer, the reversible chemical reaction forming dimers from monomers is represented as follows.
\begin{align}
M+M\rightleftharpoons D
\label{mech: simple dimer}
\end{align}
When formed \emph{in vitro}, monomers equilibrate with dimers rapidly, on the order of minutes. The chromatography experiments under consideration require on the order of hours to perform. Hence, the experimental technique is unable to resolve the reaction rates.

In the dimer exchange assay, some of the monomers are tagged with extra atoms, making them chemically distinct from the normal type, but without appreciably altering the way in which dimers form. The altered monomer is said to be of the fusion type.
\begin{myassum}The addition of MBP to LBD to create the fusion protein does not appreciably alter dimerization kinetic parameters. \label{assum: heterobind
}\end{myassum}
The two types of monomers are now labeled $M_{1}$ and $M_{2}$.
\begin{subequations}
\label{mech: dimer exchange}
\begin{align}
&M_{1}+M_{1}\rightleftharpoons D_{11}\\
&M_{2}+M_{2}\rightleftharpoons D_{22}\\
&M_{1}+M_{2}\rightleftharpoons D_{12}
\end{align}
\end{subequations}
Biochemists call $D_{11}$ and $D_{22}$ \emph{homodimers} and $D_{12}$ a \emph{heterodimer}. See Figure~\ref{fig: dimershapes}. Understanding the rates at which these dimers form is essential to an understanding of their biological actions.  The dimer exchange assay creates a slowly equilibrating system, allowing reaction rates to be more easily studied. See Figure~\ref{fig: dimerexchangedata} for a typical time course. 

The \emph{Law of Mass Action} \cite{deVries2006, Guldberg:1879fk, Scott:1994aa} states that the rate at which chemical reactions occur is proportional to the products of concentrations of the chemically reacting species. Applying this law to reaction \eqref{mech: simple dimer} yields a system of differential equations.
\begin{subequations}
\label{de: simple dimer}
\begin{align}
&\frac{dM}{dt}=-2k_{+}M^{2}+2k_{-}D \label{de: simple dimerM} \\
&\frac{dD}{dt}=k_{+}M^{2}-k_{-}D \label{de: simple dimerD}
\end{align}
\end{subequations}
The dependent variables $M(t)$ and $D(t)$ are concentrations of monomers and dimers, respectively. The coefficients $k_{+}$ and $k_{-}$ are \emph{rate parameters}. They are not rates, because their units are $1/$concentration*time and $1/$time, respectively. Rates have units of concentration$/$time. Also note the factor of 2 in equation \eqref{de: simple dimerM}. This is due to the 2:1 stoichiometric ratio of monomers to dimers.

Applying the law of mass action to \eqref{mech: dimer exchange} leads to a more complex model.
\begin{subequations}
\label{de: dimer exchange}
\begin{align}
&\frac{dM_{1}}{dt}=-2k_{1,+}M_{1}^{2}+2k_{1,-}D_{11}-k_{3,+}M_{1}M_{2}+k_{3,-}D_{12}\\
&\frac{dM_{2}}{dt}=-2k_{2,+}M_{2}^{2}+2k_{2,-}D_{22}-k_{3,+}M_{1}M_{2}+k_{3,-}D_{12}\\
&\frac{dD_{11}}{dt}=k_{1,+}M_{1}^{2}-k_{1,-}D_{11}\\
&\frac{dD_{22}}{dt}=k_{2,+}M_{2}^{2}-k_{2,-}D_{22}\\
&\frac{dD_{12}}{dt}=k_{3,+}M_{1}M_{2}-k_{3,-}D_{12}
\end{align}
\end{subequations}
The constants $k_{i,\pm}$ are the rate parameters for the six reactions in system \ref{mech: dimer exchange}. A subscript $+$ ($-$) denotes a forward (backward) rate constant. The subscript indices $i=1,2,3$ correspond to reactions \ref{mech: dimer exchange}a,b,c, respectively.

It has been implicitly assumed that concentrations are altered only by a closed system of chemical reactions. While it is certain that no source of monomers or dimers exists in the experimental apparatus, it is an outstanding question whether appreciable losses of protein are occurring by aggregation or binding to the glass of the column. Under the conditions of the experiment, proteins should be stable, but the possibility of degradation can't be completely ruled out either.
\begin{myassum}Protein does not exit the closed chemical reaction system described by Equations~\eqref{de: dimer exchange} during the dimer exchange assay. \label{assum: noloss}\end{myassum}
\noindent This assumption will be revisited in \S\ref{sec: stats}.

\subsection{Model Reduction}
System \eqref{de: dimer exchange} has five unknown concentrations. Examining experimental data from Figure~\ref{fig: dimerexchangedata} and Table~\ref{tab: dimerexchangedata} shows that biochemists don't have measurements for all five concentrations. This is due to difficulty of distinguishing between small monomer concentrations and noise in the chromatogram. System \eqref{de: dimer exchange} also contains six unknown rate constants. Again, the experimental data shows there will only be three curves to fit, which may be insufficient to accurately approximate six parameters.

\subsubsection{Conservation Laws}
By examining the simple monomer scenario \eqref{mech: simple dimer}, it is intuitively clear that monomers are never destroyed, they are simply incorporated into dimers. Each time a dimer forms, two monomers are used. So the total concentration of monomers, counting both the free and dimerized forms, is $2D+M$, and this quantity should be unchanging in time. Combining equations \eqref{de: simple dimerM} and \eqref{de: simple dimerD} confirms this.
\begin{align*}
\frac{d}{dt}\left(2D+M\right)=2\frac{dD}{dt}+\frac{dM}{dt}=2\left(k_{+}M^{2}-k_{-}D\right)-2k_{+}M^{2}+2k_{-}D=0
\end{align*}
The quantity $2D+M$ is \emph{conserved}. As a result, it is equal to its initial value.
\begin{align*}
2D(t)+M(t)=2D(t_0)+M(t_0)\equiv\alpha
\end{align*}
Using this to replace $M$ in equation \eqref{de: simple dimerD} yields
\begin{align}
\label{de: simple dimerReduced}
\frac{dD}{dt}=k_{+}\left(\alpha-2D\right)^{2}-k_{-}D \, .
\end{align}
Evidently, conservation laws enable the reduction of a system of ODE \cite{Scott:1994aa}. The trajectory of $(M,D)$ in two-dimensional concentration space is restricted to a one-dimensional \emph{stoichiometric subspace}~\cite{Feinberg:1974fk}.

The same perspective applied to the dimer exchange system \eqref{mech: dimer exchange}, reveals that each of the two types of monomers aren't destroyed, they only change form. This provides two conservation laws.
\begin{subequations}
\label{eq: cons laws}
\begin{align}
&\frac{d}{dt}\left(M_{1}+2D_{11}+D_{12}\right)=0\\
&\frac{d}{dt}\left(M_{2}+2D_{22}+D_{12}\right)=0
\end{align}
\end{subequations}
These can be rephrased as algebraic equations involving initial conditions.
\begin{subequations}
\begin{align*}
&M_{1}(t)+2D_{11}(t)+D_{12}(t)=M_{1}(t_0)+2D_{11}(t_0)+D_{12}(t_0)\equiv\alpha_{1}\\
&M_{2}(t)+2D_{22}(t)+D_{12}(t)=M_{2}(t_0)+2D_{22}(t_0)+D_{12}(t_0)\equiv\alpha_{2}
\end{align*}
\end{subequations}
When applied to system \eqref{de: dimer exchange}, these conservation laws reduce the model.
\begin{subequations}
\label{de: dimer exchange 2}
\begin{align}
&\frac{dD_{11}}{dt}=k_{1,+}(\alpha_{1}-2D_{11}-D_{12})^{2}-k_{1,-}D_{11}\\
&\frac{dD_{22}}{dt}=k_{2,+}(\alpha_{2}-2D_{22}-D_{12})^{2}-k_{2,-}D_{22}\\
&\frac{dD_{12}}{dt}=k_{3,+}(\alpha_{1}-2D_{11}-D_{12})(\alpha_{2}-2D_{22}-D_{12})-k_{3,-}D_{12} \quad .
\end{align}
\end{subequations}
Identifying two conservation laws was done intuitively. If the reaction system were more complex, intuition alone could be insufficient to identify these laws. Also, this system could allow additional conservation laws. A complete treatment requires finding others or proving that there are no others.

Though system \eqref{de: dimer exchange} is nonlinear, there is a reformulation which reveals underlying linearity. Chemical reactions are modeled by autonomous systems, $x'=F(x)$, with $F$ a nonlinear mapping from species to rates of change of species. This mapping can be decomposed, $F(x)=L\Phi(x)$, where $\Phi$ is a mapping from species to complexes and $L$ is a linear mapping from complexes to rates of change of species.
\begin{align*}
\frac{d}{dt}\begin{bmatrix}M_{1}\\M_{2}\\D_{11}\\D_{22}\\D_{12}\end{bmatrix}=
\begin{bmatrix}-2k_{1,+}&0&-k_{3,+}&2k_{1,-}&0&k_{3,-}\\
0&-2k_{2,+}&-k_{3,+}&0&2k_{2,-}&k_{3,-}\\
k_{1,+}&0&0&-k_{1,-}&0&0\\
0&k_{2,+}&0&0&-k_{2,-}&0\\
0&0&k_{3,+}&0&0&-k_{3,-}
\end{bmatrix}
\begin{bmatrix}M_{1}^{2}\\M_{2}^{2}\\M_{1}M_{2}\\D_{11}\\D_{22}\\D_{12}\end{bmatrix}
\end{align*}

\begin{mydef} Given a vector of species concentrations $x$, a conservation law for a chemical reaction system is any constant linear combination of concentrations which is unchanging in time.
\begin{align*}
0=\frac{d}{dt}\left(v_{1}x_{1}+v_{2}x_{2}+\ldots+v_{n}x_{n}\right)=\frac{d\phantom{t}}{dt}(v^{T}x)
\end{align*}
\end{mydef}
\noindent A conservation law is uniquely represented (up to scalar multiples) by the corresponding vector $v$. This definition motivates a means of identifying conservation laws.

\begin{mythm} Let $x'=L\Phi(x)$ be a chemical reaction system whose right hand side has been decomposed as described above. Then $v$ is a conservation law of the system only if $v\in\ker L^{T}$.
\end{mythm}
\begin{proof} Suppose $v$ is a conservation law for $x'=L\Phi(x)$.
\begin{align*}
(L^Tv)^T\Phi(x)=(v^{T}L)\Phi(x)=v^{T}x'=(v^{T}x)'=0
\end{align*}
\end{proof}

\noindent This necessary condition is applied to dimer exchange.
\begin{align*}
L^Tv=
\begin{bmatrix}-2k_{1,+}&0&k_{1,+}&0&0\\
0&-2k_{2,+}&0&k_{2,+}&0\\
-k_{3,+}&-k_{3,+}&0&0&k_{3,+}\\
2k_{1,-}&0&-k_{1,-}&0&0\\
0&2k_{2,-}&0&-k_{2,-}&0\\
k_{3,-}&k_{3,-}&0&0&-k_{3,-}
\end{bmatrix}
\begin{bmatrix}v_{1}\\v_{2}\\v_{3}\\v_{4}\\v_{5}\end{bmatrix}
=
\begin{bmatrix}0\\0\\0\\0\\0\end{bmatrix}
\end{align*}
The kernel of $L^T$ is given by
\begin{align*}
v=v_{1}\begin{bmatrix}1&0&2&0&1\end{bmatrix}^T+v_{2}\begin{bmatrix}0&1&0&2&1\end{bmatrix}^T \ .
\end{align*}
The conservation laws form a two dimensional space spanned by the intuited conservation laws in system~\eqref{eq: cons laws}. When analyzing large networks, or small networks where intuition is lacking, this method of identifying all possible conservation laws is effective and simple to implement.

It's notable that the conservation laws for the dimer exchange problem are independent of the rate parameters. Further decomposition of $L$ allows many interesting conclusions to be drawn about the reaction network without regard to the underlying differential equations or rate parameters. This is the subject of Chemical Reaction Network Theory~\cite{Bailey:2001vn,Feinberg:1979fr,Feinberg:1987uq,Feinberg:1974fk,Shinar:2009kx}.

One advantage of deducing conservation laws in this way is that choices for $v_{1}$ and $v_{2}$ lead to other versions of the conservation laws, ones which may be less intuitive. For example $v_{1}=v_{2}=1$ leads to a law for the conservation of the total number of monomers. While $v_{1}=1=-v_{2}$ shows that if the difference in monomer concentrations changes, it is accompanied by a two fold change in the difference between homodimer concentrations.

\subsubsection{Experimental Observations}
\label{sec: exp_obs}

As stated in \S \ref{sec: rate laws}, the addition of atoms to create fusion monomer $M_{2}$ is believed to only affect its diffusion through the chromatography column, without appreciably influencing its dimerization. This implies the following.
\begin{subequations}
\begin{align}
&k_{1,-}=k_{2,-}=k_{3,-}\ , \\
&k_{1,+}=k_{2,+}=k_{3,+}/2\ . \label{eq: kplusratio}
\end{align}
\end{subequations}
The second equality in~\eqref{eq: kplusratio} follows from a simple probability argument.

Suppose there are $N$ molecules of each of two types occupying some volume. Suppose further that a collision randomly occurs between two molecules. If the molecules are of different types, then there are $N^2$ possible collisions. If the molecules are of the same type, then there are $N(N-1)/2$ possible collisions. For large values of $N$, these numbers differ by approximately a factor of 2. Because chemical reactions are due to random collisions, and because concentration values are simply numbers of molecules scaled by volume, the second equality in~\eqref{eq: kplusratio} follows.

At the start of the experiment separate solutions of $(M_{1},D_{11})$ and $(M_{2},D_{22})$ are prepared and allowed to reach equilibrium. It's known that at equilbrium the concentrations of monomers are several orders of magnitude smaller than the dimers. These solutions are then combined, with a small number of additional monomers being freed initially.
Because the amount of free monomers can't be determined at the start of the experiment, and because their concentrations are believed to be about 1000 times lower than those of dimers, the initial concentration of monomers is assumed to be zero, \emph{i.e.}, $M_{1}(t_0)=M_{2}(t_0)=0$.
\begin{subequations}
\begin{align}
&2D_{11}(t_0)+D_{12}(t_0)+M_{1}(t_0)=2D_{11}(t_0)+D_{12}(t_0)\equiv 2d_{11}+d_{12}\\
&2D_{22}(t_0)+D_{12}(t_0)+M_{2}(t_0)=2D_{22}(t_0)+D_{12}(t_0)\equiv 2d_{22}+d_{12}
\end{align}
\end{subequations}

\begin{myassum}Initial concentrations of monomers are negligible.
\end{myassum}

The remaining initial concentrations are known. After these reductions, system \eqref{de: dimer exchange 2} now has three unknown concentrations, $\{D_{11},D_{22},D_{12}\}$, and two unknown parameters, $\{k_{+},k_{-}\}$. For simplicity of notation, the dimer concentrations are henceforth denoted by $\{x,y,z\}$ respectively.
\begin{subequations}
\label{de: dimer exchange reduced}
\begin{align}
&\frac{dx}{dt}=k_{+}(2d_{11}+d_{12}-2x-z)^{2}-k_{-}x, &&x(t_0)=d_{11},\\
&\frac{dy}{dt}=k_{+}(2d_{22}+d_{12}-2y-z)^{2}-k_{-}y, &&y(t_0)=d_{22},\\
&\frac{dz}{dt}=2k_{+}(2d_{11}+d_{12}-2x-z)(2d_{22}+d_{12}-2y-z)-k_{-}z, &&z(t_0)=d_{12}.
\end{align}
\end{subequations}

It's important to note that standard nonlinear regression techniques used by biochemists are able fit exponential models to the type of data displayed in Figure~\ref{fig: dimerexchangedata}. Due to the disparity between the time scales for association and dissociation, this regression gives estimates of the dissociation parameter $k_{-}$ only. These exponential fits agree somewhat with available experimental data on dimer concentrations. However, without experimental measurements of monomer concentrations, information on association was thought to be lost, making estimates of $k_{+}$ unattainable. The methods described in subsequent sections give the first estimates of $k^{+}$ from available chromatograms.

\section{Computational Modeling}
\label{sec: compmodel}

\subsection{Numerical ODE Methods}
Analytical solutions to nonlinear ODE are rarely possible to obtain. To compute numerical approximations of these solutions, the software package MATLAB is used~\cite{MATLAB:2015}. The types of ODE being solved with the range of parameters used and the numerical error tolerances required lead to the numerical solutions exhibiting \emph{stiffness}, that is, the dynamic step size adjustments of the explicit Runge-Kutta solver ode45 are made unnecessarily small to achieve stability. So called \emph{stiff} solvers based on Rosenbrock methods or backwards differentiation formulae are more efficient choices, \emph{e.g.}, MATLAB's ode23s and ode15s. The larger stability regions of these methods allow accuracy to be achieved with larger step sizes.~\cite{Rosenbrock:1963aa, Shampine:1979aa, Shampine:1997aa}.

\subsection{Parameter Fitting}
\label{sec: param fit}
Performing numerical simulations requires some \linebreak
knowledge of the rate parameters and initial conditions. As Murray~\cite[p. 417]{Murray:2002vn} stated, ``\ldots parameter estimates \dots are essential in any practical application of a model to a specific biological problem."

To estimate rate parameters, a method of computing the error between the empirical data and the numerical solution is needed. One common and simple choice is the sum of square errors and the corresponding root mean square error\footnote{The RMSE is related to the $l_2$ norm which is sometimes replaced with other $l_p$ norms or the cosine measure. A variety of other special purpose norms are available~\cite{Cha:2007aa,Deshpande:2013aa}.}. See~\cite{Aguilar:2015vn} for a detailed discussion of the statistical basis for least squares minimization applied to parameter estimation problems.

Let $x_i$ be the experimental value approximated by numerical solution $x(t)$ at time $t_i$. Define $\{y_i,z_i\}$ similarly. Define the \emph{sum of square errors} and the \emph{root mean square error}.
\begin{align*}
&\textrm{SSE}=\sum_{i=1}^{n}\left((x(t_i)-x_i)^2+(y(t_i)-y_i)^2+(z(t_i)-z_i)^2\right),
&&\textrm{RMSE}=\sqrt{\textrm{SSE}/3n}.
\end{align*}
To each pair of independent variables $(k_{+},k_{-})$ corresponds a value of the  dependent variable, SSE. In this way an error surface is generated. An algorithm is needed which minimizes the error, \emph{i.e.}, estimates the location of the global minimum on the error surface. Such algorithms successively choose parameter values to diminish the SSE until it reaches desired tolerances.

\begin{figure}[htbp]
	\centering 
	\setlength\figureheight{3cm} 
	\setlength\figurewidth{5cm}
	\input{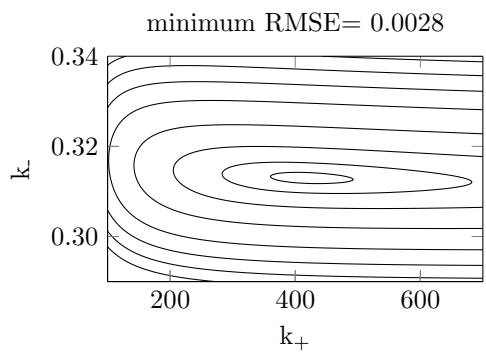}
	\caption[Error Surface]{A contour plot of the error surface for a typical dimer exchange experiment. The minimum occurs at $(k_{+},k_{-})\approx(418,0.314)$.}
	\label{fig: errorsurf}
\end{figure}

Examining Figure~\ref{fig: errorsurf} reveals that, for a typical dimer exchange assay, the local minimum on the error surface is located in a basin which is steep in the vertical direction but shallow in the horizontal direction. Indeed, in the figure $k_+$ varies over $[100,700]$ while $k_-$ varies over $[0.29,0.34]$, an aspect ratio of 12,000. When a minimum is trapped in such a narrow region, optimization algorithms which rely on gradient estimates tend to perform poorly or fail altogether. A famous example demonstrating this difficulty is the Rosenbrock Banana Function \cite{Rosenbrock:1960aa}. 

One algorithm which is effective for this type of optimization is the Nelder-Mead Simplex Method \cite{Nelder:1965aa,Wright:2012fk}. It's implemented by MATLAB's \emph{fminsearch}. Given a starting guess for the parameters, the simplex algorithm deterministically selects a succession of parameter pairs which tend towards a local minimum on the error surface.

\begin{figure}[htbp]
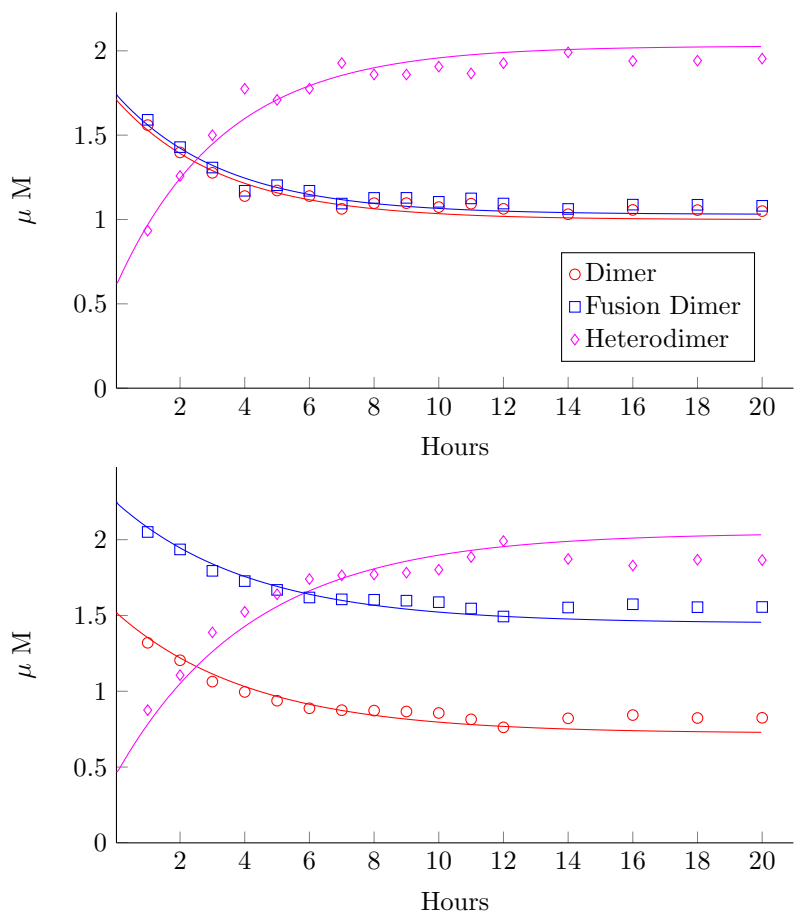

	\centering 
	\setlength\figureheight{5cm} 
	\setlength\figurewidth{9cm}
	\input{Best_Fit_3_8.tikz}
	\input{Best_Fit_3_13_b.tikz}
	\caption[Best Fit]{The best fit numerical solutions for two typical dimer exchange assays. Top: Data from Table~\ref{tab: dimerexchangedata2} leads to best fit parameters $(k_{+},k_{-})=(284,0.238)$. Bottom: Data from Table~\ref{tab: dimerexchangedata3} leads to best fit parameters $(k_{+},k_{-})=(439,0.301)$.}
	\label{fig: best fit}
\end{figure}

When multiple local minima are suspected, multi-start methods~\cite{Marti:2003vn,Solis:1981ys} or simulated annealing~\cite{Bertsimas:1993uq,Kirkpatrick:1983kx} are often used to search for the global minimum. Such methods were not applied to dimer exchange, but their frequent use in applications motivates the following brief descriptions.

Suppose it is known that a surface has multiple local minima contained in some finite domain. Starting from a single initial parameter estimate and proceeding with a minimization algorithm leads to a numerical estimate of a single local minimum. If the numerical experiment is repeated with different start values, it's possible that the same local minima, or a different one, will be found. Repeating this process with many initial parameter guesses spread over parameter space is the idea of a \emph{multi-start} method. If a visual inspection indicates the location of minima, then appropriate initial estimates are easy to obtain. However if the parameters are spread over a large multidimensional domain, such \emph{a  priori} knowledge of the locations of minima is not easily obtained. Finding all local minima may require a shotgun approach to be used, whereby numerous start values are selected with sufficient density to suggest that all possible local minima will be found. A multi-start method for a high dimensional problem can be computationally expensive.

Multi-start is an dense deterministic search of all relevant parameter space, while \emph{simulated annealing} is a random walk over the subregions of parameter space presumed likely to contain global minima. Annealing is a metallurgical technique for creating stable alloys by carefully heating and cooling a mixture of metals according to a prescribed temperature schedule. Alloys cooled rapidly have their molecules locked into a local minimum energy state that is often far from globally optimal. By the application of a designed cooling schedule, the molecules gradually come to their equilibrium locations. Heat promotes large molecular deviations to new unexplored energy configurations. An effective combination of heating and cooling allows sufficient opportunity for large molecular deviations (to find regions with possible global minima) and also for exploration of the depths of the energy minima.

Inspired by physics, simulated annealing is a minimization process where the next set of coordinates to be used in the minimization process are chosen according to the outcomes of previous simulations but with the addition of noise to mimic thermal effects. The amount of noise added at each round of coordinate selection follows a temperature schedule chosen to allow a balance between finding areas with minima (high temperature random large deviations) and exploring the depths of the minima (low temperature nearly deterministic small deviations). Multi-start and simulated annealing are frequently combined.

MATLAB's Global Optimization Toolbox enables applications of multi-start, simulated annealing, genetic algorithms, and other global minimization methods~\cite{MATLAB:optimization}.

An examination of a large portion of the error surface suggests that dimer exchange admits a single local minimum which may be approximated without global methods.

\begin{myrem} The parameter fitting method used for the dimer exchange assay was chosen for simplicity. In practice, Kalman Filtering and other methods~\cite{Kalman:1960ly,Lillacci:2010zr} are used to not only estimate parameters but to suggest appropriate mathematical models for systems which may be not be completely understood and to indicate which laboratory experiments should be performed next.
\end{myrem}

\subsection{Sensitivity Analysis}
\label{sec: sensitivity}
\subsubsection{Local Sensitivity}
It's important to understand how sensitive a system is to alterations of the parameters. There are many potential sources of error, possibly leading to variation in parameter estimates.
\vspace{1ex}
\begin{remunerate}
\item Experimental data may not be adequate, \emph{e.g.}, the nature of chromatography causes the first time point of the dimer exchange assay to be recorded at $t=1/30$ hours instead of $t=0$. This causes inaccuracy in the initial conditions.
\item Experimental data is never perfectly reproducible so that separate experiments lead to different parameter estimates.
\item Numerical solvers give approximate solutions, up to specified error tolerances.
\end{remunerate}
\vspace{1ex}
\noindent The second source of error is considered in \S\ref{sec: stats}. If a model is highly sensitive to changes in parameters, small experimental errors may lead to large deviations in the best fit parameters, even though large differences in experimental outcomes weren't observed. If a model is insensitive to changes in the parameters, then the experimental outcome may not be appreciably altered by those parameters. Both scenarios may imply that certain components of the model should be replaced, modified, or removed.

Suppose an ODE is solved multiple times for $x(t)$ with different, but similar, values of a model parameter, $k$. Small changes in $x(t)$ will result due to small changes in $k$.
\begin{align*}
\frac{\text{Change in }x}{\text{Change in }k}=\frac{\delta x}{\delta k}\approx\frac{dx}{dk}
\end{align*}
Consider the following initial value problem.
\begin{align}
&\frac{dx}{dt}=f(x,t;k), \quad x(0)=x_{0}(k). \label{eq: sensode}
\end{align}
The function $f$ and the initial data $x_{0}$ depend on a parameter $k$, \emph{e.g.}, an autocatalysis model with initial data at the concentration of half-maximal production rate, $x'=x^{2}/(x^{2}+k^2)$ with $x(0)=k$.
\begin{mydef}Let $x(t;k)$ be the solution to initial value problem~\eqref{eq: sensode}. The \emph{local sensitivity} of $x$ with respect to $k$ is defined as $y(t;k)=\frac{d x}{d k}$.
\end{mydef}

\begin{mythm}
The local sensitivity satisfies the following ODE.
\begin{align*}
\frac{d y}{d t}=\frac{\partial f}{\partial x}y+\frac{\partial f}{\partial k}
\end{align*}
\end{mythm}
\begin{proof}
By the chain rule and Clairaut's theorem,
\begin{align*}
\frac{d y}{d t}=\frac{d}{d t}\frac{d x}{d k}=\frac{d}{d k}\frac{d x}{d t}=\frac{df}{d k}=\frac{\partial f}{\partial x}y+\frac{\partial f}{\partial k} \ .
\end{align*}
\end{proof}
\noindent As an example, sensitivity analysis is applied to system \eqref{de: simple dimerReduced} with $y=\frac{d D}{d D_{0}}$.
\begin{align*}
&\frac{dD}{dt}=k_{+}\left(M_{0}+2D_{0}-2D\right)^{2}-k_{-}D, &&D(0)=D_{0},\\
&\frac{dy}{dt}=\left(-4k_{+}\left(M_{0}+2D_{0}-2D\right)-k_{-}\right)y+4k_{+}\left(M_{0}+2D_{0}-2D\right), &&y(0)=1.
\end{align*}
Local sensitivity analysis may be applied to systems of ODE with multiple parameters. Specialized numerical methods have been designed for both local and global sensitivity analysis~\cite{Gonnet:2012aa, Saltelli:2005aa}. In general, $n$ equations with $m$ parameters leads to a system of $n(m+1)$ equations for the solutions and their parameter sensitivities. See Equations~\eqref{de: dimer exchange reduced sensitivity} in Appendix~\ref{app: sensitivity} for the nine equations used in the sensitivity analysis of solution $\{x,y,z\}$ with respect to parameters $\{k_{+},k_{-}\}$.

Of interest is how a relative change in the solution would arise in response to a relative change in a parameter. That is, $k$ will be perturbed, the change in $x$ will be found, and ratios of relative changes will be computed.

\begin{mydef} The \emph{relative local sensitivity} of solution $x$ with respect to parameter $k$ is $ky/x$.
\end{mydef}
\noindent The motivation of this definition is intuitive.
\begin{align*}
\frac{\text{Relative change in }x}{\text{Relative change in }k}=\frac{\frac{\delta x}{x}}{\frac{\delta k}{k}}=\frac{k}{x}\frac{\delta x}{\delta k}\approx \frac{k}{x} y
\end{align*}
An alternative expression for the relative sensitivity is, by the chain rule $\frac{d \ln x}{d \ln k}$.  See~\cite{Shinar:2009kx} for an application.

\begin{figure}[htbp]
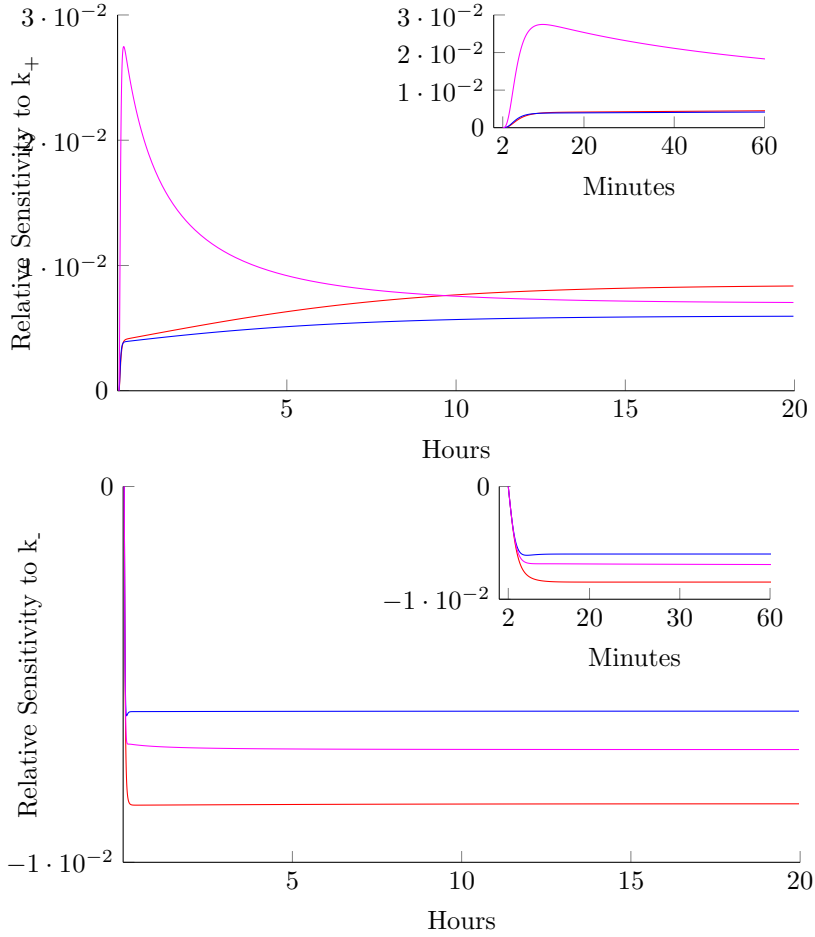

	\centering 
	\setlength\figureheight{5cm} 
	\setlength\figurewidth{9cm}
	\input{Sensitivity_kp3_fix.tikz}
	\input{Sensitivity_km3_fix.tikz}
	\caption[Sensitivity]{The data from Table~\ref{tab: dimerexchangedata2} leads to best fit parameters $[k_{+}, k_{-}]= [284, 0.238]$. See Figure~\ref{fig: best fit}, top. For these values, relative sensitivities of $\{x,y,z\}$ were computed with respect to $k_{+}$ (top) and $k_{-}$ (bottom). Top: At hour 20, the relative sensitivities to $k_+$ in decreasing order are for dimer (red), heterodimer (purple), and fusion dimer (blue). Bottom: at hour 20, the relative sensitivities to $k_-$ in decreasing order are for fusion dimer (blue), heterodimer (purple), and dimer (red).}
	\label{fig: sensitivity}
\end{figure}

Figure~\ref{fig: sensitivity} shows relative local sensitivities of solutions $\{x,y,z\}$ to changes in parameters $\{k_{+},k_{-}\}$. This figure reveals that relative sensitivity of $z$ to $k_+$ peaks during the first hour. Unfortunately, during this initial phase only one data point was recorded by the experimentalists. Limitations of the experimental protocol make recording more data in this interval challenging. Nonetheless, this new observation suggests that more accurate estimates of $k_+$ may be found by restricting simulation and data fitting to the early part of the experiment.

The figure was produced by numerically solving the ODE for the local sensitivities using standard MATLAB ODE solvers. Users of MATLAB's SimBiology tool may compute sensitivities automatically without the need for explicitly stating the governing ODE~\cite{MATLAB:simbiology}.

\subsubsection{Global Sensitivity}

Local sensitivities are computed by means of derivatives of outcome variables with respect to single parameters, \emph{e.g.}, $dx/dk_{+}$. As such, these sensitivities are most informative when changes in parameters and outcomes are sufficiently small so as to be well approximated by infinitesimals and when parameters are sufficiently independent so that changes in outcomes due to each parameter may be examined separately. Many biological experiments show large deviations in, and nonlinear interactions between, parameters. A global approach to the study of sensitivity is often appropriate.

While local sensitivity can be analyzed by solving additional ODE, global sensitivity has been defined in many ways and sophisticated statistical tools are often required to analyze it~\cite{Saltelli:2004kx,Turani:1990aa}. Such analyses typically proceed in three stages. Firstly, knowledge and intuition of the possible ranges for the values of the $n$ parameters are used to predetermine the parameter set of interest. Next, a finite subset of n-tuples are sampled from this set and the computational simulation is run $n$ times using those parameters. Finally, the $n$ simulation outcomes are statistically analyzed to determine how alterations to each parameter affected the outcomes.

The Fourier Amplitude Sensitivity Test (FAST) is a method for studying global sensitivity~\cite{Cukier:1973fk,Cukier:1973fk2}. In its simplest form, FAST proceeds in three stages. First, the rate constants are varied parametrically at different frequencies.
\begin{align*}
&k_+(s)=k_+^0\sin(\omega_+ s), \quad k_-(s)=k_-^0\sin(\omega_- s).
\end{align*}
Here $k_\pm^0$ are representative values of $k_\pm$, $s$ is a parameter defined on $[-\pi,\pi]$, and $\omega_{\pm}$ are integers. Then, for a specified time and for all $s$, an outcome variable, \emph{e.g.},  $x(s)$, is computed and its Fourier coefficients are found.
\begin{align}
&x(s)=\frac{1}{2}A_{0}+\sum_{j=1}^{\infty}A_{j}\cos(js)+B_{j}\sin(js), \nonumber \\
&A_j=\frac{1}{2\pi}\int_{-\pi}^\pi x(s)\cos(j s)\, ds, \quad B_j=\frac{1}{2\pi}\int_{-\pi}^\pi x(s)\sin(j s)\, ds. \label{eq: fouriercoef}
\end{align}
Parseval's theorem shows that the sum of squares of these coefficients is proportional to the variance of $x(s)$.
\begin{align*}
\Var(x)=\frac{1}{2\pi}\int_{-\pi}^{\pi}x(s)^{2}\,ds-\left(\frac{1}{2\pi}\int_{-\pi}^{\pi}x(s)\,ds\right)^{2}=\frac{1}{2}\sum_{j=1}^{\infty}\left(A_{j}^{2}+B_{j}^{2}\right)
\end{align*}
Finally, by summing only those squares of coefficients corresponding to frequencies which are multiples of the $\omega$'s, the variance due to the changes in a particular rate constant can be discerned. This motivates the computation of a \emph{sensitivity index}, which is the ratio of the variance due to one parameter to the total variance. For example,
\begin{align*}
S_{x,k_+}=\frac{\sum_{i=1}^{\infty} \big(A_{i\omega_+}^2+B_{i\omega_+}^2\big)}{ \sum_{j=1}^{\infty} \left(A_j^2+B_j^2\right)}.
\end{align*}
By definition, sensitivity indices are elements of $[0,1]$. The selection of $G_\pm$, $\omega_\pm$, and a finite set of $s$ values to accurately approximate the integrals~\eqref{eq: fouriercoef} are subtle issues addressed elsewhere~\cite{Cukier:1973fk,Marino:2008fk,Saltelli:1999uq}.

For any time point of interest, the sensitivity indices determine which portion of the variance can be attributed to each parameter. By repeating the analysis for different times, global sensitivity indices may be plotted in a manor similar to the relative local sensitivities of Figure~\ref{fig: sensitivity}.

\begin{figure}[htbp]
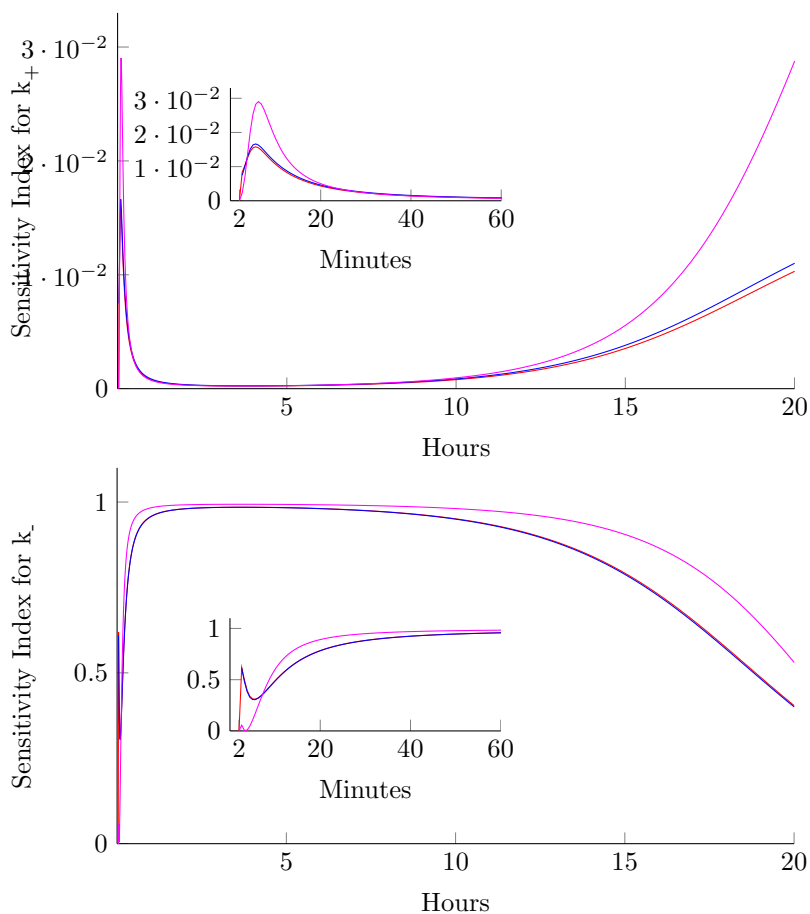

	\centering 
	\setlength\figureheight{5cm} 
	\setlength\figurewidth{9cm}
	\input{GlobalSensitivity_kp_fix.tikz}
	\input{GlobalSensitivity_km_fix.tikz}
	\caption[Global Sensitivity]{Global sensitivity analysis using FAST with parameter ranges $k_{+}\in[200,350]$ and $k_{-}\in[0.2,0.35]$. The sensitivity index for a parameter is the ratio of variance in an outcome variable at the parameter's frequency (and harmonics) to the total variance in the outcome variable. Top: At hour 20, the $k_+$ sensitivity indices in decreasing order are for heterodimer (purple), fusion dimer (blue), and dimer (red). Bottom: At hour 20, the $k_-$ sensitivity indices in decreasing order are for heterodimer (purple), dimer (red), and fusion dimer (blue). The latter two of which are nearly visually indistinguishable.}
	\label{fig: GSA_efast}
\end{figure}

The interested reader is referred to~\cite{Marino:2008fk} and~\cite{Pianosi:2015uq}, which provide background on global sensitivity and whose authors supply open source software. Dimer exchange was studied using MATLAB routines from the former. These routines implement the authors' extended version of the algorithm, eFAST~\cite{software:sensitivity}. Users of MATLAB's \emph{Simulink} have access to a suit of tools appropriate for the study of global sensitivity~\cite{MATLAB:sensitivity}.

Applying eFAST to dimer exchange results in Figure~\ref{fig: GSA_efast}. As with relative local sensitivity, global sensitivity to alterations in $k_+$ peaks during the first hour of the experiment, highlighting the possible utility of a protocol which would allow additional data to be collected during the initial phase. These results have motivated new experimental design which, if successful, may enable data collection over shorter time intervals.

Comparing the vertical scale of the two plots in Figure~\ref{fig: GSA_efast} suggests much greater sensitivity of model outcomes to large deviations in $k_-$. The relative insensitivity of outcomes to large deviations in $k_+$ poses difficulties when estimating best fit parameters. Variance in parameter estimates is addressed in \S\ref{sec: stats}.

\section{Data Modeling}
\label{sec: stats}

The effect of variations in the experimental data on the computed optimal rate constants can be studied using data analytic methods. The sensitivity of computational outcomes to parameter values was highlighted in \S\ref{sec: sensitivity}. Fitting data to a single experiment gives rate constants which describe that single experiment. By examining a set of $n$ experiments, a set of $n$ pairs of rate constant estimates can be found. If the experimental outcomes are consistent, and if the model and numerical simulations are accurate, then the computed rate constants should be consistent across all experiments. It's necessary to give analytical meaning to this vague expectation of \emph{consistency}.

The data methods employed below fall into two categories, 1) graphical methods which give qualitative information about the rate constants and 2) analytical methods which give quantitative information. 

\begin{figure}[htbp]
	\centering
	\setlength\figureheight{3cm} 
	\setlength\figurewidth{4.25cm}
%
%
%
%
\begin{tikzpicture}

\begin{axis}[%
width=\figurewidth,
height=\figureheight,
scale only axis,
xmin=0,
xmax=12,
xtick={2,5,8,11},
xticklabels={$10^2$,$10^5$,$10^8$,$10^{11}$},
xlabel={$k_+$},
ymin=0,
ymax=0.5,
ytick={0.1, 0.2, 0.3, 0.4},
ylabel={$k_-$},
axis x line*=bottom,
axis y line*=left
]
\addplot[scatter,only marks,scatter src=explicit,scatter/classes={
1={mark=*,mark options={},black!10!gray},%
2={mark=triangle*,mark options={},blue},%
3={mark=pentagon*,mark options={},red},%
4={mark=square*,mark options={},black!30!green}
}] plot coordinates{
(1.16545207283193,0.29962) [1]
(11.118727550427,0.26409) [3]
(1.15390628513887,0.30599) [1]
(1.90134387489848,0.3159) [1]
(8.52200058318427,0.33097) [3]
(10.1364351704584,0.23218) [3]
(10.110455024461,0.21014) [3]
(9.80656659536504,0.32438) [3]
(10.0230465840755,0.23839) [3]
(1.25085895459929,0.26094) [1]
(10.0063804585497,0.2521) [3]
(0.676446677310898,0.33171) [1]
(11.173477643453,0.20178) [3]
(11.0770043267934,0.26513) [3]
(1.39684420528474,0.14684) [2]
(2.16244491429998,0.17865) [2]
(9.59863732804622,0.11385) [4]
(1.20387562327733,0.25195) [1]
};

\end{axis}
\end{tikzpicture}%
\hspace{0.5cm}
%
%
%
%
\begin{tikzpicture}

\begin{axis}[%
width=\figurewidth,
height=\figureheight,
scale only axis,
xmin=0,
xmax=5,
xtick={1,2,3,4},
xticklabels={$10^{1}$,$10^{2}$,$10^3$,$10^4$},
xlabel={$k_+$},
ymin=0,
ymax=0.5,
axis x line*=bottom,
hide y axis
]
\addplot[scatter,only marks,scatter src=explicit,scatter/classes={
1={mark=square*,mark options={},black!30!green},%
2={mark=*,mark options={},black!10!gray},%
3={mark=triangle*,mark options={},blue},%
4={mark=pentagon*,mark options={},red}
}] plot coordinates{
(2.38637409908548,0.29534) [2]
(2.62708921305274,0.23754) [2]
(2.40447450573985,0.29786) [2]
(2.89275667525507,0.30885) [2]
(2.97636373679429,0.3045) [2]
(2.88258714470419,0.21566) [2]
(3.07532793416326,0.19376) [2]
(2.64277108920975,0.3005) [2]
(3.68238879074338,0.22305) [1]
(2.83206800780853,0.25934) [2]
(2.45344065929356,0.23845) [2]
(4.39701832704231,0.31366) [4]
(3.04292973334316,0.15091) [2]
(2.5127510825101,0.18068) [2]
(3.82297835488452,0.14617) [3]
(3.03869962302062,0.17279) [2]
(3.89765457741316,0.10982) [3]
(2.42452229708663,0.24403) [2]
};

\end{axis}
\end{tikzpicture}%
\hspace{0.5cm}
	
	\caption[Parameter Clustering]{Left: Cluster analysis applied to raw data. Right: Cluster analysis applied to processed data from Table~\ref{tab: bestfitparams} in the appendix.}
	\label{fig: cluster}
\end{figure}
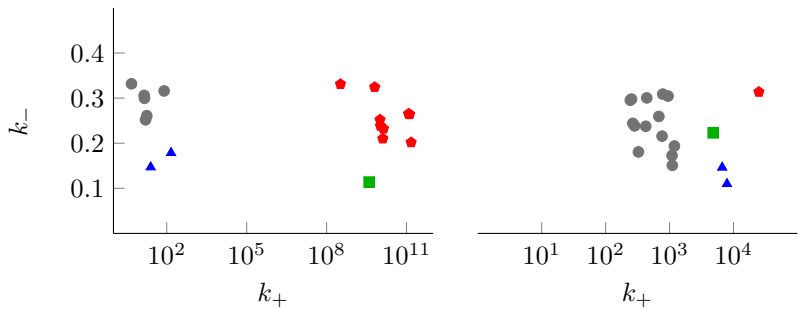

The left of Figure~\ref{fig: cluster} shows a scatter plot of the optimal rate constants for the raw data sets from 18 dimer exchange assays. It is clear that $k_+$ estimates vary greatly. Indeed, it should be noted that $k_{+}$ values in excess of $10^{8}$ cause the numerical ODE solver to fail to achieve desired error tolerances. So these anomalously large values of $k_{+}$ are dubious.

The raw data used to obtain these parameter estimates exhibits artifacts which suggest to experimentalists that some loss of protein is occurring during the assay. A corrective processing of the concentration data was used to account for these losses. At each time step, the data was renormalized so that total protein of LBD type and of Fusion type are invariant in time. The resulting parameter fits to processed data are shown in on the right of Figure~\ref{fig: cluster}. The new estimates for $k_{+}$ are within the limitations of the numerical ODE solver.

Cluster Analysis \cite{Bezdek:1981aa, Datta:2003aa, Day:1985aa, MacQueen:1967aa, Tan:2006aa} may be used to qualify what is meant by the terms \emph{outlier} or \emph{atypical}. Once identified, these outliers should be examined to determine the cause of their anomalous nature. MATLAB's \emph{clusterdata} function uses a hierarchical clustering method, though it can employ other methods such as k-means and Gaussian mixtures models.

The left plot of Figure~\ref{fig: cluster} shows a cluster of experiments (red pentagons) which do not fit the mathematical model because of suspected protein loss, highlighting the need for an improved model or corrective processing of the data. The right plot demonstrates the effect of processing to account for protein loss. Cluster analysis reveals that 14 of 18 parameter pairs are in the same cluster (grey circles).

\begin{mydef} Let $\{p_{k}\}$ be the set of data points, $\{c_{k}\}$ be the set of clusters the data has been separated into, and $\mu(p_i,p_j)$ be a measure of distance between two data points. Given a point $p_{i}$ from cluster $c_{k_{i}}$ containing $|c_{k_i}|$ elements, the \emph{silhouette width}\footnote{This is the original definition, given by Rousseeuw~\cite{Rousseeuw:1987fk}. Alternate definitions, including the one used by MATLAB, assign $S=1$ to singleton clusters.} of $p_{i}$  is defined by
\begin{align*}
S(p_{i})=
\left\{\begin{array}{cc}
\frac{B_{i}-A_{i}}{\max(A_{i},B_{i})} & \text{if } |c_{k_i}|\neq1 \\
0 &  \text{if } |c_{k_i}|=1
\end{array}\right.
\end{align*}
where $A_{i}$ is the mean distance from $p_{i}$ to all points in the same cluster and $B_{i}$ is the minimum mean distance from $p_{i}$ to all points in other clusters.
\begin{align*}
&A_{i}=\frac{1}{| c_{k_i}|-1} \sum_{ p_{j}\in c_{k_i} } \mu(p_{i},p_{j})\, , &&B_{i}=\min_{k\neq k_{i}}\bigg(\frac{1}{|c_{k}|}\sum_{p_{j}\in c_{k}}\mu(p_{i},p_{j})\bigg)\, .
\end{align*}
\end{mydef}

\noindent Note that $S(p)\in[-1,1]$ for all $p$. A silhouette value near $+1$ indicates that a point is well matched to its assigned cluster and poorly matched to the others. A negative silhouette value indicates that a point may have been assigned to the wrong cluster.

To create Figure~\ref{fig: cluster}, silhouette widths were computed for various numbers of clusters. Choosing four clusters gave an optimal value for the average silhouette width. The reader is encouraged to investigate MATLAB's \emph{silhouette} and \emph{evalclusters} functions. For readers familiar with the R statistical programming language, 30 methods for choosing the number of clusters have been incorporated into the package NbClust~\cite{Charrad:2014aa}.

Variation in the computed parameter estimates can arise from many sources. Alterations may have been made to the experimental protocol or the post processing of the data. These changes may have been made intentionally by the experimenters or unintentionally due to the difficulty of controlling experimental conditions. Even given consistent data, the mathematical model based on this data may be incomplete and unable to capture phenomena exhibited in all experiments.

Using raw data, the computational model of dimer exchange failed to fit the experimental model. This is explained by Assumption~\ref{assum: noloss}, which stated that losses of protein would be ignored when forming the mathematical model. This is in conflict with the experimentalist's suspicion of protein loss.  Corrective processing of the data was used by experimentalists in an attempt to compensate for the protein loss hypothesis. An alternative would be to reformulate the model to include protein loss. Data processing and model alterations will be discussed further in \S\ref{sec: conclusions}.

Visual inspection can be deceiving, leading to dubious subjective conclusions. Changing the scales of the axes in Figure~\ref{fig: cluster} could make the rate constants look more or less associated. The clustering method used here also requires subjective specification of the number of clusters to search for. Although sophisticated methods of cluster number selection exist, ambiguity and subjectivity persist.

Objective analytical methods should be used to compliment intuition gained from graphical displays and cluster analysis.  After performing a cluster analysis, the 14 pairs of rate constants in the main cluster (grey circles) were used to compute the statistical means, standard deviations, and coefficients of variation.

\begin{align*}
&\mu_{k_+}\approx628\, , &&\sigma_{k_+}\approx352\, , &&CV_{k_+}=\sigma_{k_+}/\mu_{k_+}\approx1.78\, , \\
&\mu_{k_-}\approx0.243\, , &&\sigma_{k_-}\approx0.0540\, , &&CV_{k_-}=\sigma_{k_-}/\mu_{k_-}\approx0.223\,.
\end{align*}
The standard deviations suggest that $k_+$ estimates may be much more variable than those for $k_-$, consistent with conclusions drawn from a visual inspection of Figure~\ref{fig: cluster}. However, the scales for the rate constants are different by roughly three orders of magnitude.

To compare such dissimilar data sets, the standard deviation as a percentage of the mean is often used. These \emph{coefficients of variation} \cite{Pearson:1896aa,Reed:2002ve} show the forward rate constants to deviate from the mean about eight times as much as the backward constants. The statistical measures used here provide a superficial glimpse into the dimer exchange data. Much deeper analyses of parameter estimates are common~\cite{Aguilar:2015vn}. Future work aims to reduce the variance in $k_+$ by collecting more data from the first hour of the assay, to address the peak parameter sensitivities described in \S\ref{sec: sensitivity}. 

\section{Discussion}
\label{sec: conclusions}
Given the preponderance of wet-lab biology data, mathematical methods are needed to analyze and give meaning to the data. The process employed above is an example of one such type of analysis.

The first stage is the acquisition of experimental data. Information from experimenters and trusted mathematical techniques are then used to form an analytical model of the mechanism underlying the experimental protocol. The model is reduced by utilizing biological knowledge and analytical techniques. The model is then simulated in order to fit parameters to the empirical data, using sensitivity analysis as a guide toward the most relevant parts of the experiment. Data analytic methods may then detect atypical experiments and allow statistical inferences of the trustworthiness of parameter estimates

Accurate models and parameter estimates should not be considered the ultimate goal of this process. To complete the cycle of experimentation and modeling, conclusions drawn from the data should be used to inform the experimental model.

Indeed, the outliers may highlight breaches of experimental protocol. Large variance may point towards insufficient data collection or an ineffective experimental technique. Analytical models with poor fit to the empirical data may indicate that the model is inadequate or that the underlying biology is misunderstood.

The present study has revealed the need for biochemists to process the data to account for protein loss. It also motivates the mathematician to rebuild models to account for losses (aggregation, non-specific binding, degradation, etc.) so that such data processing isn't required. This is the subject of future collaborative work.

The models considered here were presented to biochemists studying estrogen receptors. The outcomes strengthened their hypothesis that protein aggregation was occurring prior to injection of the mixture into the chromatography column. This hypothesis has been incorporated into a more complex mathematical model not presented here. The researchers are``extremely interested" in learning the outcome of this new model, as protein aggregation is an important area of contemporary research~\cite{Brandt:2014kx}.

Nonlinear regression techniques used by biochemists are able fit exponential models to the type of data displayed in Figure~\ref{fig: dimerexchangedata}. Due to the disparity between the time scales for association and dissociation, this regression gives estimates of the dissociation parameter $k_{-}$ but not of $k_{+}$. These exponential fits agree with available experimental data on dimer concentrations. However, without experimental measurements of monomer concentrations, information on association was assumed to be lost.

The analytical methods demonstrated above give the first estimates of $k^{+}$. \linebreak
Though monomer concentrations aren't discernible, their presence evidently leaves a shadow in the dimer concentration data. The models and methodology above are the first to shed light on these shadows. Motivated by sensitivity analysis of heterodimer concentrations with respect to $k_{+}$, one goal of future experiments is to find ways to collect data over shorter time intervals.

The realm of collaboration between mathematicians and biologists extends beyond the analysis of data to support existing hypotheses. The goal of collaboration is to add mathematical analysis to the set of tools available to biologists, that is, to enable mathematics to be a new type of laboratory equipment.


\begin{appendices}
\appendixpage

\section{Experimental Data}
\label{app: data}
In the tables of this section, the quantities $D_{11}$, $D_{22}$, and $D_{12}$ are the experimentally measured concentrations of LBD homodimer, fusion homodimer, and heterodimer, respectively. The units of concentration are micromolar. Time is recorded in hours. Limitations of the chromatography process make it impossible to record data before two minutes, \emph{i.e.}, $1/30$ hour.

To produce this data, raw chromatograms were processed to account for protein loss. At each time step, the data was renormalized so that total protein of LBD type and of Fusion type are invariant in time. All data is from unpublished work from the Brandt Lab~\cite{Brandt:2008a,Brandt:2007a}.

\begin{table}[H]
\footnotesize
\caption{Data from the dimer exchange assay plotted in Figure~\ref{fig: dimerexchangedata}.}
\begin{center}
\renewcommand{\arraystretch}{1.2}
\renewcommand{\tabcolsep}{2pt}
\begin{tabular}{|c|c|c|c|c|c|c|c|c|c|c|c|c|c|c|c|}
\hline
t&1/30&1&2&3&4&5&7&9&11&13&15&18&23&29&35 \\
\hline
$D_{11}$&1.88&1.65&1.46&1.33&1.23&1.18&1.10&1.06&1.03&1.02&1.00&1.00&1.01&0.99&0.99\\
\hline
$D_{22}$&2.08&1.86&1.67&1.53&1.44&1.39&1.30&1.26&1.24&1.22&1.20&1.19&1.21&1.19&1.19\\
\hline
$D_{12}$&0.16&0.62&1.00&1.27&1.46&1.57&1.72&1.81&1.86&1.90&1.93&1.93&1.91&1.96&1.96\\
\hline
\end{tabular}
\end{center}
\label{tab: dimerexchangedata}
\end{table}%

\begin{table}[H]
\footnotesize
\caption{Data from the dimer exchange assay plotted in Figure~\ref{fig: best fit}.}
\begin{center}
\renewcommand{\arraystretch}{1.2}
\renewcommand{\tabcolsep}{2pt}
\begin{tabular}{|c|c|c|c|c|c|c|c|c|c|c|c|c|c|c|c|c|c|}
\hline
t&1/30&1&2&3&4&5&6&7&8&9&10&11&12&14&16&18&20 \\
\hline
$D_{11}$&1.71&1.56&1.40&1.28&1.14&1.17&1.14&1.06&1.10&1.10&1.07&1.09&1.06&1.03&1.06&1.06&1.05\\
\hline
$D_{22}$&1.75&1.59&1.43&1.31&1.17&1.20&1.17&1.09&1.13&1.13&1.10&1.13&1.09&1.06&1.09&1.09&1.08\\
\hline
$D_{12}$&0.62&0.93&1.26&1.50&1.78&1.71&1.78&1.93&1.86&1.86&1.91&1.87&1.93&1.99&1.94&1.94&1.95\\
\hline
\end{tabular}
\end{center}
\label{tab: dimerexchangedata2}
\end{table}

\begin{table}[H]
\footnotesize
\caption{Data from the dimer exchange assay plotted in Figure~\ref{fig: best fit}.}
\begin{center}
\renewcommand{\arraystretch}{1.2}
\renewcommand{\tabcolsep}{2pt}
\begin{tabular}{|c|c|c|c|c|c|c|c|c|c|c|c|c|c|c|c|c|c|}
\hline
t&1/30&1&2&3&4&5&6&7&8&9&10&11&12&14&16&18&20 \\
\hline
$D_{11}$&1.52&1.32&1.20&1.06&1.00&0.94&0.89&0.88&0.87&0.87&0.86&0.81&0.76&0.82&0.84&0.82&0.83\\
\hline
$D_{22}$&2.25&2.05&1.94&1.79&1.73&1.67&1.62&1.61&1.60&1.60&1.59&1.55&1.49&1.55&1.57&1.55&1.56\\
\hline
$D_{12}$&0.47&0.88&1.11&1.39&1.52&1.54&1.74&1.76&1.77&1.78&1.80&1.89&1.99&1.87&1.83&1.87&1.87\\
\hline
\end{tabular}
\end{center}
\label{tab: dimerexchangedata3}
\end{table}%

\section{Local Sensitivities}
\label{app: sensitivity}

Define local sensitivities by
\begin{align*}
&x_1=\frac{dx}{dk\mathrlap{_+}}\phantom{_+} , &&x_2=\frac{dx}{dk\mathrlap{_-}}\phantom{_-} , &&y_1=\frac{dy}{dk\mathrlap{_+}}\phantom{_+} , &&y_2=\frac{dy}{dk\mathrlap{_-}}\phantom{_-} , &&z_1=\frac{dz}{dk\mathrlap{_+}}\phantom{_+} , &&z_2=\frac{dz}{dk\mathrlap{_-}}\phantom{_-} .
\end{align*}
The differential equations governing local sensitivity with respect to $\{k_+,k_-\}$ are
\begin{subequations}
\label{de: dimer exchange reduced sensitivity}
\begin{align}
&\frac{dx_1}{dt}=(-4k_{+}(2d_{1}+d_{12}-2x-z)-k_{-})x_1+(2d_{1}+d_{12}-2x-z)^{2},\\
&\frac{dx_2}{dt}=(-4k_{+}(2d_{1}+d_{12}-2x-z)-k_{-})x_2-x,\\
&\frac{dy_1}{dt}=(-4k_{+}(2d_{2}+d_{12}-2y-z)-k_{-})y_1+(2d_{2}+d_{12}-2y-z)^{2},\\
&\frac{dy_2}{dt}=(-4k_{+}(2d_{2}+d_{12}-2y-z)-k_{-})y_2-y,\\
&\frac{dz_1}{dt}=(-4k_{+}(d_1+d_{2}+d_{12}-x-y-z)-k_{-})z_1\nonumber \\
&\phantom{\frac{dz_1}{dt}=} +2(2d_{1}+d_{12}-2x-z)(2d_{2}-2y-z),\\
&\frac{dz_2}{dt}=(-4k_{+}(d_1+d_{2}+d_{12}-x-y-z)-k_{-})z_2-z,
\end{align}
with $x_i(t_0)=0$, $y_i(t_0)=0$, and $z_i(t_0)=0$.
\end{subequations}

\section{Best Fit Parameters for 18 Experiments}
Data fitting routines were used to compute pairs of best fit parameters for 18 sets of processed dimer exchange data. The parameters $k_{+}$ and $k_{-}$ have units of $1/$micromolar*hours and $1/$hours, respectively.
\begin{table}[H]
\footnotesize
\caption{Parameter estimates for processed data corresponding to Figure~\ref{fig: cluster}.}
\begin{center}
\renewcommand{\arraystretch}{1.2}
\renewcommand{\tabcolsep}{1.2pt}
\begin{tabular}{|l|c|c|c|c|c|c|c|c|c|c|c|c|c|c|c|c|c|c|}
\hline
$k_{+}\times10^{-2}$&2.39&2.63&2.40&2.89&2.98&2.88&3.08&2.64&3.68&2.83&2.45&4.40&3.04&2.51&3.82&3.04&3.90&2.42 \\
\hline
$k_{-}\times10^{1}$&2.95&2.38&2.98&3.09&3.05&2.16&1.94&3.01&2.23&2.59&2.38&3.14&1.51&1.81&1.46&1.73&1.10&2.44 \\
\hline
\end{tabular}
\end{center}
\label{tab: bestfitparams}
\end{table}%

\section{MATLAB Algorithms}
The primary algorithms used to simulate models and process data for the dimer exchange assay are listed and described below.

\begin{description}
\item[ode15s] This implicit solver is useful for solving ODE systems with widely varying time scales. It is very commonly used in mathematical biology. Compare to the explicit solver \emph{ode45}, which is more useful for problems which are not stiff. Use \emph{odeset} to specify conditions such as error tolerances.
\item[fminsearch] This minimization method often succeeds when gradient-based methods, such as \emph{lsqcurvefit}, fail. Use \emph{optimset} to specify conditions such as error tolerances.
\item[clusterdata] This is an agglomerative hierarchical clustering routine packaging together several MATLAB functions. Its high level nature makes it simple to use. The low level details may be explored through its options or the related functions \emph{linkage} and \emph{cluster}.
\item[silhouette] The silhouette value for each point is a measure of how similar that point is to points in its own cluster, when compared to points in other clusters. Silhouettes can be useful for determining an appropriate number of clusters to seek.
\item[evalclusters] A routine which, given a clustering algorithm, determines an appropriate choice for the number of clusters by attempting to optimize silhouette widths, gap statistics, or other criteria.
\end{description}

\end{appendices}

\bibliographystyle{siam}

\end{document}